\documentstyle[12pt,a4,epsf]{article}
 \renewcommand{\baselinestretch}{1.5}
\newtheorem{rema}{\hskip\parindent\sc Remark}
\newtheorem{theo}{\hskip\parindent\sc Theorem}
\newtheorem{coll}{\hskip\parindent\sc Corollary}

 \makeatletter
 \@addtoreset{equation}{section}
 \makeatother

\begin{document}
\catcode`@=11
\newskip\plaincentering \plaincentering=0pt plus 1000pt minus 1000pt
\def\plainLet@{\relax\iffalse{\fi\let\\=\cr\iffalse}\fi}
\def\plainvspace@{\def\vspace##1{\noalign{\vskip##1}}}
\newif\iftagsleft@
\tagsleft@true
\def\TagsOnRight{\global\tagsleft@false}
\TagsOnRight
\def\tag#1$${\iftagsleft@\leqno\else\eqno\fi
 \hbox{\def\pagebreak{\global\postdisplaypenalty-\@M}%
 \nopagebreak{\global\postdisplaypenalty\@M}\rm(#1\unskip)}%
  $$\postdisplaypenalty\z@\ignorespaces}
\interdisplaylinepenalty=\@M
\def\plainallowdisplaybreak@{\def\allowdisplaybreak{\noalign{\allowbreak}}}
\def\plaindisplaybreak@{\def\displaybreak{\noalign{\break}}}
\def\align#1\endalign{\def\tag{&}\plainvspace@\plainallowdisplaybreak@\plaindisplaybreak@
  \iftagsleft@\plainlalign@#1\endalign\else
   \plainralign@#1\endalign\fi}
\def\plainralign@#1\endalign{\displ@y\plainLet@\tabskip\plaincentering\halign to\displaywidth
     {\hfil$\displaystyle{##}$\tabskip=\z@&$\displaystyle{{}##}$\hfil
       \tabskip=\plaincentering&\llap{\hbox{(\rm##\unskip)}}\tabskip\z@\crcr
             #1\crcr}}
\def\plainlalign@
 #1\endalign{\displ@y\plainLet@\tabskip\plaincentering\halign to \displaywidth
   {\hfil$\displaystyle{##}$\tabskip=\z@&$\displaystyle{{}##}$\hfil
   \tabskip=\plaincentering&\kern-\displaywidth
        \rlap{\hbox{(\rm##\unskip)}}\tabskip=\displaywidth\crcr
               #1\crcr}}
\catcode`@=12

\def\sect{\section{\leftline{\large\bf}}}
\thispagestyle{empty}
\vspace*{-13mm}

\begin{center}
{\large\bf ESTIMATION FOR SINGLE-INDEX MIXED MODELS WITH
LONGITUDINAL DATA}
\\[0.5cm]
{\large Zhen Pang${^1}$  and  Liugen Xue${^2}$} \\[0.4cm]
{\it $^1$Division of Mathematical Sciences, School of Mathematical Sciences, Nanyang Technological University, Singapore}\\
{\it $^2$College of Applied Sciences, Beijing University of
Technology, Beijing, China }
\end{center}
\vspace{0.2cm}

\begin{center}
\renewcommand{\baselinestretch}{1.5}
\parbox{0.8\hsize}
{~~~~\small In this paper, we consider a single-index mixed model
with longitudinal data. A new set of estimating equations is proposed to estimate the single-index coefficient.
The link function is estimated by using the local linear
smoothing. Asymptotic normality is established for the proposed
estimators. Also, the estimator of the link function achieves
optimal convergence rates; and the estimators of variance
components have root-$n$ consistency. These results facilitate the
construction of confidence regions/intervals and hypothesis
testing for the parameters of interest. Some simulations and an
application to real data are included. }
\end{center}
\vspace{0.2cm}

\footnotetext{Zhen Pang's research was supported by one research grant from the 
Nanyang Technological University.}

 \footnotetext{Liugen Xue's research was supported by the National
Natural Science Foundation of China (10871013), the Beijing
Natural Science Foundation (1102009) and the PHR(IHLB).}



\newpage

\section{Introduction }

\hskip\parindent
 Consider the single-index mixed model
\begin{eqnarray}
Y_{ij} = g(X_{ij}^T\beta_0) + \alpha_i + \varepsilon_{ij}, \ \
i=1,\ldots,n,~ j=1,\ldots,m,
  \label{(1.1)}
\end{eqnarray}
where $\alpha_i$ and $\varepsilon_{ij}$ are independent mean zero
random variables with variances $\sigma_\alpha^2>0$ and
$\sigma_\varepsilon^2>0$, respectively, $g(\cdot)$ is an unknown
link function, and $\beta_0$ is a $p\times1$ vector of unknown
parameters. For the sake of identifiability, it is often assumed
that $\|\beta_0\|=1$ and the first nonzero component of $\beta_0$
is positive, where $\|\cdot\|$ denotes the Euclidean metric. Let
$Y_i=(Y_{i1},\ldots,Y_{im})^T$, $X_i=(X_{i1},\ldots,X_{im})^T$,
$\varepsilon_i=(\varepsilon_{i1},\ldots,\varepsilon_{im})^T$ and
$G(X_i\beta_0)=(g(X_{i1}^T\beta_0),\ldots,g(X_{im}^T\beta_0))^T$.
The model implies that the $Y_i$ are independent with
$E(Y_i|X_i)=G(X_i\beta_0)$ and ${\rm
cov}(Y_i|X_i)=V=\sigma_\alpha^2\textbf{1}_m\textbf{1}_m^T +
\sigma_\varepsilon^2{\rm I_m}$, where $\textbf{1}_m$ is an $m\times
1$ vector of ones and ${\rm I_m}$ is the $m \times m$ identity matrix.

We address the general problem of estimating the parameter $\beta_0$, the
function $g(\cdot)$, and the
variance components $\sigma_\alpha^2$ and $\sigma_\varepsilon^2$ simultaneously when $m$ is fixed. We will
show in Section 3 that the variance components
$\sigma_\alpha^2$ and $\sigma_\varepsilon^2$ can be estimated at the
parametric rate $O_P(n^{-1/2})$ which allows us to treat them as known when we derive the theoretical results for $\beta_0$ and $g(\cdot)$ in Sections 2
and 3.

The single-index model is an important tool in multivariate
nonparametric regression, which can avoid the
so-called ``curse of dimensionality" by searching a univariate index of the multivariate covariate $X$ to capture important
features of high-dimensional data. The
single-index model has been applied in a variety of fields, such as discrete
choice analysis in econometrics and dose-response models in
biometrics (H\"{a}rdle et al. 1993). In the cross-sectional data,
many authors have studied the statistical inference problem of the
single-index model, and reported many results, for example,
Li (1991), Ichimura (1993), Zhu and Ng (1995), Xia and Li (1999),
Naik and Tsai (2000), Hristache, Juditsky and Spokoiny (2001), Xia
et al. (2002), Stute and Zhu (2005), Xia (2006), and Xue and Zhu
(2006). Meanwhile, the estimation problem of the partially linear
single-index model has been widely addressed as well by
Carroll et al. (1997), Yu and Ruppert (2002), Xia and H\"{a}rdle
(2006), Zhu and Xue (2006), Wang et al. (2010) and others. These reported methods have been proven to be useful and effective for the independent data. On the other hand, to our knowledge, the method to treat correlated data, which are commonly seen in econometrics and biometrics, is lacking in literature. In this paper, such models will be developed and reported.

Longitudinal data are perhaps the most well-known type of correlated data. There are already extensive literature on the generalized linear, nonparametric and semiparametric mixed models for longitudinal data, see, for example, Zeger and Diggle (1994), Jiang (1998), Zhang, et al.
(1998), Jiang (1999), Ruckstuhl, Welsh and Carroll (2000), Jiang
and Zhang (2001), Jiang, Jia and Chen (2001), Ke and Wang (2001),
Cai, Cheng and Wei (2002), Wu and Zhang (2002), Liang, Wu and
Carroll (2003), Zhang and Lin (2003), Gu and Ma (2005), Hall and
Maiti (2006), Jiang (2006) and Field, Pang and Welsh (2008), among others. However, literature on the applications of single-index
models for longitudinal/panel data is limited. Honor\'{a} and Kyriazidou (2000) and Carro (2007) proposed some
estimating methods for dynamic panel data discrete choice models.
Bai et al. (2009) studied the single-index models for longitudinal
data, where they proposed a procedure to estimate the single-index
component and the link function based on the combination of the
penalized splines and quadratic inference functions. Liang and Zeger (1986) proposed an extension of the generalized linear
models to the analysis of longitudinal data. They introduced the
generalized estimating equations (GEE) that gave consistent
estimates of the regression parameters and their variance under
mild assumptions on the time dependence. The GEE were derived
without specifying the joint distribution of a subject's
observations yet they reduced to the score equations for
multivariate Gaussian outcomes. In this paper, we apply the idea
of GEE to the single-index mixed models with longitudinal data. To
estimate the single-index coefficient $\beta_0$, we propose a new set of
estimating equations which take the constraint $\|\beta_0\|=1$
into account. The estimator based on these estimating equations outperform previous ones, as summarized below. First, our estimation
procedure does not specify a form for both the distribution of
random effect and the joint distribution of the repeated
measurements. Second, we introduce estimating equations that give
the root-$n$ consistent estimate of $\beta_0$ under week
assumptions on the joint distribution. Third, we construct the root-$n$ consistent estimates of the
variance components $\sigma_\varepsilon^2$ and $\sigma_\alpha^2$.
It allows us to consider the construction of confidence regions
and hypothesis testing for $\beta_0$. Lastly, we also obtain
the asymptotic normality and the uniform convergence rate of the
estimator of $g(\cdot)$. Our algorithm is numerically fast and stable.

The rest of the paper is organized as following. In Section 2, we
elaborate on the methodology. Section 3 presents the asymptotic
properties for all proposed estimators. Section 4 reports the
results of simulation studies and one real example. The proofs of the
main theorems are relegated to the Appendix.

\section{Estimation method}

\subsection{Estimations of the parametric and nonparametric components}

 \hskip\parindent
 If $g$ were known, we could estimate $\beta_0$ by minimizing
$$
R_n(\beta) \equiv
\frac{1}{n}\sum_{i=1}^n\{Y_i-G(X_i\beta)\}^TW(X_i\beta)V^{-1}\{Y_i-G(X_i\beta)\}
$$
for $\beta$ with $\|\beta\|=1$, where $W(X_i\beta)={\rm
diag}\{w(X_{i1}\beta),\ldots,w(X_{im}\beta)\}$, and $w(\cdot)$ is
a bounded weight function with a bounded support ${\cal U}_w$,
which is introduced to control the boundary effect. For simplicity
and convenience, we assume that $dw(u)/du=0$. Especially, we can
take $w(\cdot)=I_{[-a,a]}(\cdot)$, for some constant $a>0$. This
is a restricted least squares problem. We now use the constraint
$\|\beta_0\|=1$ to transfer the restricted least squares to
the unrestricted least squares, which makes it possible to search for
the solution of the estimating equations over a restricted region
in the Euclidean space $R^{p-1}$. For this, we need to calculate
the derivative of $g(X_{ij}^T\beta)$ at point $\beta_0$.
Note that $\|\beta_0\|=1$ means that the true value $\beta_0$ is the
boundary point of the unit sphere. The function $g(X_{ij}^T\beta)$
does not have the derivative at point $\beta_0$. For this, we
suggest the popularly used delete-one-component method (Wang et
al., 2010). The detail is as follows. Without loss of generality,
we may assume that the true parameter $\beta_0$ has a positive
component (otherwise, consider $-\beta_0$), say $\beta_{0r}>0$ for
$\beta_0=(\beta_{01},\ldots,\beta_{0p})^T$ and $1\leq r\leq p$.
For $\beta=(\beta_1,\ldots,\beta_p)^T$, let
$\beta^{(r)}=(\beta_1,\ldots,\beta_{r-1},\beta_{r+1},\ldots,\beta_p)^T$
be a $p-1$ dimensional parameter vector after removing the $r$th
component $\beta_r$ in $\beta$. Then the true parameter
$\beta_0^{(r)}$ must satisfy the constraint $\|\beta_0^{(r)}\|<
1$, and $\beta$ is infinitely differentiable in a neighborhood of
$\beta_0^{(r)}$. The {\rm Jacobian} matrix of $\beta$ with respect
to $\beta^{(r)}$ is defined as
\begin{eqnarray}
{J}_{\beta^{(r)}}=\frac{\partial\beta}{\partial{\beta^{(r)}}}=(\gamma_1,\ldots,
\gamma_p)^T,
 \label{(2.1)}
\end{eqnarray}
where $\gamma_s$ $(1\leq s\leq p, s\neq r)$ is a $p-1$ dimensional
unit vector with $s$th component 1, and
$\gamma_r=-(1-\|\beta^{(r)}\|^2)^{-1/2}\beta^{(r)}$. Let
$X_{ij}=(X_{ij1},\ldots,X_{ijp})^T$ and
$X_{ij}^{(r)}=(X_{ij1},\ldots,X_{ij(r-1)},X_{ij(r+1)},\ldots,X_{ijp})^T$.
Then we have
$X_{ij}^T\beta=X_{ij}^{(r)T}\beta^{(r)}+(1-\|\beta^{(r)}\|^2)^{1/2}X_{ijr}$,
which is a function of $\beta^{(r)}$. When $g$ is known, we can
obtain an estimator of $\beta_0^{(r)}$ by solving
\begin{eqnarray}
Q_n(G,\beta^{(r)})\equiv\frac{1}{n}\sum_{i=1}^nJ_{\beta^{(r)}}^TX_i^TG'_\Delta(X_i\beta)W(X_i\beta)V^{-1}\{Y_i-G(X_i\beta)\}=0
 \label{(2.2)}
\end{eqnarray}
for $\beta^{(r)}$, where $G'_\Delta(X_i\beta)={\rm
diag}\{g'(X_{i1}\beta),\ldots,g'(X_{im}\beta)\}$. An iteratively
reweighted least squares algorithm is widely used for solving this
system of equations. Given a current estimate
$\tilde{\beta}_0^{(r)}$ with $\|\tilde{\beta}_0^{(r)}\|=1$,
compute
$$
\tilde{\beta}^{(r)}=\tilde{\beta}_0^{(r)} +
B_n^{-1}(G,\tilde{\beta}_0^{(r)})Q_n(G,\tilde{\beta}_0^{(r)})
$$
and
$\tilde{\beta}^{(r)}=\tilde{\beta}^{(r)}/\|\tilde{\beta}^{(r)}\|$,
where
\begin{eqnarray*}
B_n(G,\beta^{(r)})\equiv\frac{1}{n}\sum_{i=1}^nJ_{\beta^{(r)}}^TX_i^TG_\Delta^{'2}(X_i\beta)W(X_i\beta)V^{-1}
X_iJ_{\beta^{(r)}}.
\end{eqnarray*}
This iteratively reweighted least squares algorithm solves
(\ref{(2.2)}) and is identical to the Fisher's method of scoring
version of the Newton-Raphson algorithm for solving these
estimating equations. Using $\|\beta_0\|=1$ and $\|\beta\|=1$, we
can prove
$$
{\beta}-{\beta}_0=J_{{\beta}_0^{(r)}}({\beta}^{(r)}-{\beta}^{(r)})
+ O_P(n^{-1}).
$$
Thus, we can obtain an iterative formula for estimating $\beta$
when $g$ is known, that is
\begin{eqnarray}
\hat{\beta}^*=\tilde{\beta} +
J_{\tilde{\beta}_0^{(r)}}B_n^{-1}(G,\tilde{\beta})Q_n(G,\tilde{\beta})
 \label{(2.3)}
\end{eqnarray}
and $\hat{\beta}^*=\hat{\beta}^*/\|\hat{\beta}^*\|$, where the
initial value of $\beta_0$, say $\|\tilde{\beta}_0\|=1$, can be
obtain by fitting the linear model. Then, set
$\tilde{\beta}=\hat{\beta}^*$ and iterate until convergence.

Since we assume that the link function $g$ is unknown, it must be
estimated. Given an initial estimate $\tilde{\beta}_0$ of
$\beta_0$, we can easily compute a nonparametric estimates
$\hat{g}$ and $\hat{g}'$ of $g$ and $g'(u)$. We employ the local
linear smoother (Fan and Gijbels, 1996) to obtain estimators of
the link function $g$ and its derivative $g'$. Specifically, for a
kernel function $K(\cdot)$ on the real set $R^1$ and a bandwidth
sequence $h=h_n$ tending to 0, define
$K_h(\cdot)=h^{-1}K(\cdot/h)$. For a fixed $\beta$, the local
linear smoother aims at minimizing the weighted sum of squares
\begin{eqnarray}
\sum_{i=1}^n\sum_{j=1}^m\{Y_{ij}-d_0-d_1(X_{ij}^T\beta_0-u)\}^2K_h(X_{ij}^T\beta_0-u)
 \label{(2.4)}
\end{eqnarray}
with respect to the parameters $d_\nu$, $\nu=0, 1$. Let
$\hat{d}_0$ and $\hat{d}_1$ be the solutions to the weighted least
squares problem (\ref{(2.4)}). The local linear estimators for
$g(u)$ and $g'(u)$ are defined as $\hat{g}(u;\beta_0)=\hat{d}_0$
and $\hat{g}'(u;\beta_0)=\hat{d}_1$ at the fixed point $\beta_0$.
It follows from the theory of least squares that
\begin{eqnarray}
\left(\hat{g}(u;\beta_0),h\hat{g}'(u;\beta_0)\right)^T
=S_n^{-1}(u;\beta_0)\xi_n(u;\beta_0),
 \label{(2.5)}
\end{eqnarray}
where
$$
S_n(u;\beta_0)=\left(
 \begin{array}{cc}
 S_{n,0}(u;\beta_0) & S_{n,1}(u;\beta_0) \\
 S_{n,1}(u;\beta_0) & S_{n,2}(u;\beta_0)
 \end{array} \right)
$$
and
$$
\xi_n(u;\beta_0)=\left(\xi_{n,0}(u;\beta_0),~
\xi_{n,1}(u;\beta_0)\right)^T
$$
with
\begin{eqnarray}
S_{n,l}(u;\beta_0)=\frac{1}{n}\sum_{i=1}^n\sum_{j=1}^m\left(\frac{X_{ij}^T\beta_0-u}{h}\right)^lK_h(X_{ij}^T\beta_0-u)
 \label{(2.6)}
\end{eqnarray}
and
\begin{eqnarray}
\xi_{n,l}(u;\beta_0)=\frac{1}{n}\sum_{i=1}^n\sum_{j=1}^mY_{ij}\left(\frac{X_{ij}^T\beta_0-u}{h}\right)^lK_h(X_{ij}^T\beta_0-u)
 \label{(2.7)}
\end{eqnarray}
for $l=0,1,2$.

The estimator $\hat{g}$ is called pooled estimator in existing
literatures, for example Lin and Carroll (2000), Ruckstuhl,
Welsh and Carroll (2000), and Xue (2010). As pointed out in these
literatures, the simple pooled estimator which ignores the
dependence structure performs very well asymptotically.

When $g$ is unknown, we can also obtain an estimator of
$\beta_0^{(r)}$ by solving the estimating equations
$Q_n(\hat{G},\beta^{(r)})=0$, where
$\hat{G}(X_i\beta_0)=(\hat{g}(X_{i1}^T\beta_0),\ldots,\hat{g}(X_{im}^T\beta_0))^T$
and $\hat{G}'_\Delta(X_i\beta)={\rm
diag}\{\hat{g}'(X_{i1}\beta),\ldots,\hat{g}'(X_{im}\beta)\}$ for
$i=1,\ldots,n$. We propose the use of an alternating algorithm;
first estimating $\beta_0$, and then the link function $g$,
repeating these until certain criterion is met. Given $\hat{g}$ and
$\hat{g}'$, we use the scoring algorithm (\ref{(2.3)}) to estimate
$\beta_0$, that is
\begin{eqnarray}
\hat{\beta}=\tilde{\beta} +
J_{\tilde{\beta}^{(r)}}B_n^{-1}(\hat{G},\tilde{\beta})Q_n(\hat{G},\tilde{\beta})
 \label{(2.8)}
\end{eqnarray}
and $\hat{\beta}=\hat{\beta}/\|\hat{\beta}\|$; given the estimate
of $\beta_0$, we used the pooled estimate (\ref{(2.5)}) to get a
new estimate of the link function $g$.

With $\hat{\beta}$, the final estimator of $g$ can be defined by
$\hat{g}^*(u)=\hat{g}\big(u;\hat{\beta}\big)$. The asymptotic
result for the estimate of link function $g$ follows from Theorems
\ref{theo1} and \ref{theo2}, and the result for the estimate of
parameter $\beta_0$ is established in Theorem \ref{theo3}.

\begin{rema}
\label{rema1}\rm\ We consider a homoscedastic model of (1.1).
While the estimation procedure can be extended to heteroscedastic
errors. In addition, the single-index assumption in (\ref{(1.1)})
can be readly extended to multiple indices through Sliced Inverse Regression (SIR) or its
variants, but the estimation of the multivariate link function $g$
would encounter the curse of high dimensionality. In many applications, since no more
than three indices will be needed, the
approach in this paper can indeed  be extended in practice to
multiple indices.
\end{rema}

\subsection{Estimations of the variance components}

 \hskip\parindent
 The estimation of the nonparametric component and the asymptotic variances of
all the estimators depends on the variance components, thus we need
to exhibit consistent estimators of the variance components.

A useful approach to estimate the variance components is to
pretend that the residuals are of mean zero and have the covariance matrix
same as if $g(\cdot)$ were known. If we assume that the random
effects $\alpha_i$ and the errors $\varepsilon_{ij}$ are
Gaussianly distributed, then the observation $Y_{i}$ have
independent $N(G(X_i\beta_0),V)$ distributions. Replacing
$g(\cdot)$ and $\beta_0$ with their estimators $\hat{g}(\cdot)$
and $\hat{\beta}$, respectively, the Gaussian ``likelihood" for
$\sigma_\varepsilon^2$ and $\sigma_\alpha^2$ can be written as
\begin{eqnarray*}
&&-n(m-1)\log(\sigma_\varepsilon^2)-n\log(\sigma_\varepsilon^2+m\sigma_\alpha^2)
 -\frac{m}{\sigma_\varepsilon^2+m\sigma_\alpha^2}\sum_{i=1}^n(\bar{Y}_i-\bar{\hat{g}}_i)^2
 \\
 & & \qquad - \frac{1}{\sigma_\varepsilon^2}\sum_{i=1}^n\sum_{j=1}^m\left\{Y_{ij}-\hat{g}(X_{ij}^T\hat{\beta})-(\bar{Y}_i-\bar{\hat{g}}_i)\right\}^2,
\end{eqnarray*}
where $\bar{Y}_i=m^{-1}\sum_{j=1}^mY_{ij}$ and
$\bar{\hat{g}}_i=m^{-1}\sum_{j=1}^m\hat{g}(X_{ij}^T\hat{\beta})$.
This "likelihood" is maximized at
\begin{eqnarray}
\hat{\sigma}_\varepsilon^2
 &\!\!\! = &\!\!\! \frac{1}{n(m-1)}\sum_{i=1}^n\sum_{j=1}^m\left\{Y_{ij}-\hat{g}(X_{ij}^T\hat{\beta})-(\bar{Y}_i-\bar{\hat{g}}_i)\right\}^2,
  \label{(2.9)} \\
\hat{\sigma}_\alpha^2
  &\!\!\! = &\!\!\! \frac{1}{n}\sum_{i=1}^n(\bar{Y}_i-\bar{\hat{g}}_i)^2 - \hat{\sigma}_\varepsilon^2/m,
 \label{(2.10)}
\end{eqnarray}
when $\hat{\sigma}_\alpha^2>0$, and at $\hat{\sigma}_\alpha^2=0$
and
\begin{eqnarray}
\hat{\sigma}_\varepsilon^2
 =\frac{1}{nm}\sum_{i=1}^n\sum_{j=1}^m\left\{Y_{ij}-\hat{g}(X_{ij}^T\hat{\beta})\right\}^2,
 \label{(2.11)}
\end{eqnarray}
otherwise. It can be shown that the resulting estimators have the
same convergence rate as if $g(\cdot)$ and $\beta_0$ actually were
known. The result will be given in next section.

Alternatively, we can abandon the "likelihood" and employ a method of
moments device to get the estimators (2.9)--(2.11). We can also
adjust for the loss of degrees of freedom due to estimating
$g(\cdot)$, and obtain the estimators of $\sigma_\varepsilon^2$
and $\sigma_\alpha^2$. The details can be found in Ruckstuhl, et
al. (2000).

\section{Main Results}

 \hskip\parindent
We now study the asymptotic behavior of the estimators for the nonparametric component
$g$ as well as the parametric components $\beta_0$, $\sigma_\alpha^2$ and $\sigma_\varepsilon^2$. We first list the following regularity conditions:

{\parindent=0pt
\def\toto#1#2{\rightline{\hbox to 0.85cm{#1\hss}
\parbox[t]{15.5cm}{#2}}}

\toto{(C1)}{The joint density of
$(X_{i1}^T\beta,\ldots,X_{im}^T\beta)^T$ exists, the marginal
density $f_j(u)$ of $X_{ij}^T\beta$ and the joint density
$f_{j_1j_2}(u,s)$ of $(X_{ij_1}^T\beta, X_{ij_2}^T\beta)$, for any
$j_1\neq j_2$, are continuously differentiable for $u_0\in{\cal
U}_w$ and $(u_0,s_0)\in{\cal U}_w\times{\cal U}_w$, respectively,
and there exists a $j$ such that $f_j(u)$ is bounded away from 0,
uniformly for $u\in{\cal U}_w$ and $\beta$ near $\beta_0$, where
${\cal U}_w$ is the support of $w(u)$. }
 \vskip 10pt

\toto{(C2)}{The function $g(u)$ has two bounded and continuous
derivatives, and $g_{2r}(u)$ satisfies a Lipschitz condition of
order 1 on ${\cal U}_w$, where $g_{2r}(u)$ is the $r$th component
of $g_{2}(u)$, and $g_{2}(u)=E(X_{ij}|X_{ij}^T\beta_0=u)$, $1\leq
r\leq p$. }
 \vskip 10pt

\toto{(C3)}{The kernel $K(\cdot)$ is a bounded and symmetric
probability density function with bounded support, and satisfies
the Lipschitz condition of order 1 and $\int\!u^2K(u)du\neq 0$.}
 \vskip 10pt

\toto{(C4)}{There exists an $r=\max\{4,s\}$ such that
$E(|X_{ij}|^{r})<\infty$, $E(|\alpha_i|^{r})<\infty$ and
$E(|\varepsilon_{ij}|^{r})<\infty$, and for some
$\epsilon<2-s^{-1}$ such that $n^{2\epsilon-1}h\rightarrow\infty$,
~~$i=1,\ldots, j=1\ldots,m$. }
 \vskip 10pt

\toto{(C5)}{$nh^3/\log(1/h)\rightarrow\infty$ and
$nh^4\rightarrow0$ as $n\rightarrow\infty$. }
 \vskip 10pt

\toto{(C6)}{${B}=E\Big[{J}_{\beta_0^{(r)}}^TX_1^T
 G^{'2}(X_1\beta_0)W(X_1\beta_0)V^{-1}X_1{J}_{\beta_0^{(r)}}\Big]$
is a positive definite matrix. }
 }

\begin{rema}\rm\
Condition {\rm (C1)} ensures that the denominators of
$\hat{g}(u;\beta)$ and $\hat{g}'(u;\beta)$ are, with high
probability, bounded away from 0 for $t\in{\cal U}_w$ and $\beta$
near $\beta_0$. (C2) is the standard smoothness condition. {\rm (C3)}
is the usual assumption for second-order kernels. {\rm (C4)} is a
necessary condition for the asymptotic normality of an estimator.
{\rm (C5)} is the usual condition for bandwidth. {\rm (C6)} ensures
that the limiting variances for the estimator $\hat{\beta}$ exist.
\end{rema}

Let ${\cal B}_n=\{\beta\in{\cal B}: \|\beta-\beta_0\|\leq
c_1n^{-1/2}\}$ for some positive constant $c_1$. The definition is
motivated by the fact that, since we anticipate that $\hat{\beta}$
is root-$n$ consistent, we should look for a solution of the
equations $Q_n(\hat{g},\beta^{(r)})=0$ which involves
$\beta^{(r)}$ distant from $\beta_0^{(r)}$ by order $n^{-1/2}$.
Similar restriction was also made by H\"ardle, Hall and Ichimura
(1993) and Xia and Li (1999). Denote $\mu_l=\int\!u^lK(u)du$ and
$\nu_l=\int\!K^l(u)du$, $l=1,2$.

The following theorems state the asymptotic behavior of the
estimators proposed in Section 2. We first give the uniform
convergence rates for the estimators $\hat{g}$ and $\hat{g}'$
respectively.

\

\begin{theo}\label{theo1}
~Suppose that conditions {\rm (C1)--(C4)} hold. Then
$$
  \sup_{u\in{\cal U}_w,\beta\in{\cal B}_n}\big|\hat{g}(u;\beta)-g(u)\big|
   = O_P\left((nh/\log n)^{-1/2} + h^2\right)
$$
and
$$
 \sup_{u\in{\cal U}_w,\beta\in{\cal B}_n}\big|\hat{g}'(u;\beta)-g'(u)\big|
   = O_P\left((nh^3/\log n)^{-1/2} + h\right).
$$
\end{theo}

\

The following Theorem~\ref{theo2} shows the asymptotic normality
of estimator $\hat{g}$.

\

\begin{theo}\label{theo2}
~Suppose that conditions {\rm (C1)--(C4)} hold. If $nh^5=O(1)$,
then for any $u\in{\cal U}_w$ and $\tilde{\beta}$ such that
$\|\tilde{\beta}-\beta_0\|=O_P\big( n^{-1/2}\big)$, we have
$$
\sqrt{nh}\{\hat{g}(u;\tilde{\beta})-g(u)-b(u)\}
\stackrel{D}{\longrightarrow}N(0,\sigma^2(u)).
$$
where $b(u)=(1/2)h^2\mu_2g''(u)$, and
$\sigma^2(u)=(\sigma_\alpha^2+\sigma_\varepsilon^2)\nu_0/\sum_{j=1}^mf_j(u)$.

If further assume that $nh^5\rightarrow0$, then
$$
\sqrt{nh}\{\hat{g}(u;\tilde{\beta})-g(u)\}
\stackrel{D}{\longrightarrow}N(0,\sigma^2(u)).
$$
\end{theo}

\

In Theorems \ref{theo1} and \ref{theo2}, when we start with
$\sqrt{n}$-consistent estimator for $\beta_0$, $\hat{g}$ has
uniform convergence rate and asymptotic normality. Numerous
examples of $\sqrt{n}$-consistent estimators already exist in the
literature. For instance, Hall (1989) showed that one can obtain a
$\sqrt{n}$-consistent estimator for $\beta_0$ using projection
pursuit regression. Under the linearity condition that is slightly
weaker than elliptical symmetry of $X$, Li (1991), Hsing and
Carroll (1992) and Zhu and Ng (1995) proved that SIR, proposed by
Li (1991), leads to a $\sqrt{n}$-consistent estimator of
$\beta_0$. Xia et al. (2002) proposed the minimum
average variance estimation (MAVE) and  Xia (2006) proposed a refined
version of MAVE, and both methods can provide
$\sqrt{n}$-consistent estimators for the single-index $\beta_0$.

\

\begin{theo}\label{theo3}
~Suppose that conditions {\rm (C1)--(C6)} hold. If the $r$th
component of $\beta_0$ is positive, then
$$
\sqrt{n}\big(\hat{\beta}-\beta_0\big)\stackrel{D}{\longrightarrow}
N\left(0,~{J}_{\beta_0^{(r)}}{B}^{-1}{A}{B}^{-1}{J}_{\beta_0^{(r)}}^T\right),
$$
where
 ${A}=E\Big[{J}_{\beta_0^{(r)}}^T\{X_1-G_1(X_1\beta_0)\}^T
 G^{'2}(X_1\beta_0)W^2(X_1\beta_0)V^{-1}\{X_1-G_1(X_1\beta_0)\}{J}_{\beta_0^{(r)}}\Big]$
with
$G_1(X_1\beta_0)=(g_1(X_{11}\beta_0),\ldots,g_1(X_{1m}\beta_0))^T$
and $g_1(u)=E(X_{1j}|X_{1j}^T\beta_0=u)$, and ${B}$ is defined in
condition {\rm(C6)}.
\end{theo}

\

From Theorems~\ref{theo2} and ~\ref{theo3}, we can obtain the
following corollary~\ref{coll1}.

\

\begin{coll}\label{coll1}
~Suppose that conditions {\rm (C1)--(C6)} hold. Then, for any
$u\in{\cal U}_w$,
$$
\sqrt{nh}\{\hat{g}^*(u)-g(u)\}
\stackrel{D}{\longrightarrow}N(0,\sigma^2(u)).
$$
\end{coll}
where $\sigma^2(u)$ is defined in Theorem~\ref{theo2}.

\

From Theorem~\ref{theo3}, we obtain an asymptotic result regarding
the angle between $\hat \beta$ and $\beta_0$, which can be  used
to study issues of sufficient dimension reduction.

\

\begin{coll}\label{coll2} ~Suppose that the conditions of Theorem~\ref{theo3}
hold. Then
$$
 |\hat{\beta}^T\beta_0|-1=O_P\big(n^{-1/2}\big),
$$
where $|\hat{\beta}^T\beta_0|$ is the absolute inner product.
Their inner product represents the cosine of the angle between the
two directions.
\end{coll}

\

The following theorem provides the convergence rates of the
estimators of $\sigma_\alpha^2$ and $\sigma_\varepsilon^2$,
respectively.

\

\begin{theo}\label{theo4}
~Suppose that conditions {\rm(C1)--(C6)} hold. Then
$$
 \hat{\sigma}_\varepsilon^2-\sigma_\varepsilon^2 = O_P\left(n^{-1/2}\right),
$$
$$
 \hat{\sigma}_\alpha^2-\sigma_\alpha^2 = O_P\left(n^{-1/2}\right).
$$
\end{theo}

\

To construct confidence regions for $\beta_0$, a plug-in estimator
of the limiting variance of $\hat\beta$ is needed. We define the
following estimators $\hat{B}$ and $\hat{A}$ of $B$ and ${A}$,
respectively, by $\hat{B}=B_n(\hat{g},\hat{\beta})$ and
\begin{eqnarray*}
\hat{A}=\frac{1}{n}\sum_{i=1}^nJ_{\hat{\beta}^{(r)}}^T\{X_i-\hat{G}_1(X_i\hat{\beta};\hat{\beta})\}^T\hat{G}_\Delta^{'2}(X_i\hat{\beta};\hat{\beta})W^2(X_i\hat{\beta})\hat{V}^{-1}
\{X_i-\hat{G}_1(X_i\hat{\beta};\hat{\beta})\}J_{\hat{\beta}^{(r)}},
\end{eqnarray*}
where
$\hat{V}=\hat{\sigma}_\alpha^2\textbf{1}_m\textbf{1}_m^T+\hat{\sigma}_\varepsilon^2{\rm
I}$, $\hat{G}_\Delta^{'}(X_i\hat{\beta};\hat{\beta})={\rm
diag}\{\hat{g}'(X_{i1}^T\hat{\beta};\hat{\beta}),\ldots,\hat{g}'(X_{im}^T\hat{\beta};\hat{\beta})\}$,
$\hat{G}_1(X_i\hat{\beta};\hat{\beta})=(\hat{g}_1(X_{i1}^T\hat{\beta};\hat{\beta}),\ldots,\hat{g}_1(X_{im}^T\hat{\beta};\hat{\beta}))^T$
with
$\hat{g}_1(u;\hat{\beta})=\sum_{i=1}^n\sum_{j=1}^mW_{nij}(u;\hat{\beta})X_{ij}$,
which is the estimator of $g_1(u)=E(X_{ij}|X_{ij}^T\beta_0=u)$,
\begin{eqnarray*}
W_{nij}(u;\hat{\beta})=\frac{n^{-1}K_h(X_{ij}^T\hat{\beta}-u)\{S_{n,2}(u;\hat{\beta})-(X_{ij}^T\hat{\beta}-u)S_{n,1}(u;\hat{\beta})\}}{S_{n,0}(u;\hat{\beta})S_{n,2}(u;\hat{\beta})-S_{n,1}^2(u;\hat{\beta})},
\end{eqnarray*}
and $S_{n,2}(u;\hat{\beta})$ is defined in (\ref{(2.6)}). It is
easy to prove that
${J}_{\hat{\beta}^{(r)}}\stackrel{P}{\longrightarrow}{J}_{\beta_0^{(r)}}$,
$\hat{B}\stackrel{P}{\longrightarrow}{B}$ and
$\hat{A}\stackrel{P}{\longrightarrow}{A}$. Then for any $p \times
l$ matrix ${H}$ of full rank with $l<p$, Theorem \ref{theo2} implies
that
$$
  \left(n^{-1}{H}^T{J}_{\hat{\beta}^{(r)}}\hat{B}^{-1}\hat{A}\hat{B}^{-1}{J}_{\hat{\beta}^{(r)}}^T{H}\right)^{-1/2}{H}^T(\hat{\beta}-\beta_0)
  \stackrel{D}{\longrightarrow}N(0,{I}_l).
$$
We use Theorem 10.2d in Arnold (1981) to obtain the following
limiting distribution.

\

\begin{theo}\label{theo5}
~Suppose that the conditions
of Theorem~\ref{theo2} hold. Then
$$
  (\hat{\beta}-\beta_0)^T{H}\left(n^{-1}{H}^T{J}_{\hat{\beta}^{(r)}}
  \hat{B}^{-1}\hat{A}\hat{B}^{-1}
  {J}_{\hat{\beta}^{(r)}}^T{H}\right)^{-1}
  {H}^T(\hat{\beta}-\beta_0)
  \stackrel{D}{\longrightarrow}\chi_l^2.
 $$
\end{theo}
Theorem~\ref{theo5} can be used to construct the large sample
confidence region or interval for the parameter $\beta_0$.

Applying Corollary \ref{coll1}, we can construct pointwise
confidence interval for $g(u_0)$ at a fixed point $u_0\in{\cal
U}_w$. However, we need to use the plug-in estimators for the
asymptotic bias and covariance. Obviously, the asymptotic bias and
covariance of $\hat{g}(u_0)$ are dependent on
$\sigma_\varepsilon^2$, $\sigma_\alpha^2$ and $f_j(u_0)$.
$\sigma_\varepsilon^2$ and $\sigma_\alpha^2$ have been estimated in
(\ref{(2.9)}) and (\ref{(2.10)}). The estimator of $f_j(u_0)$,
$j=1,\ldots,m$, is defined by
$$
\hat{f}_j(u_0)= \frac{1}{nh}\sum_{i=1}^nK((X_{ij}-u_0)/h).
$$
Thus, we can derive $\hat\sigma^2(u_0)$ by replacing $f_j(u_0)$,
$\sigma_\varepsilon^2$ and $\sigma_\alpha^2$ by their consistent
estimators $\hat{f}_j(u_0)$, $\hat\sigma_\varepsilon^2$ and
$\hat\sigma_\alpha^2$ respectively. Therefore, $\hat{\sigma}^2(u_0)$ is
a consistent estimator of $\sigma^2(u_0)$. By Corollary \ref{coll1},
we have
$$
\sqrt{nh}\{\hat{g}(u_0;\hat{\beta})-g(u_0)\}/\hat{\sigma}(u_0)
\stackrel{D}{\longrightarrow}N(0,1)
$$
Using above result, we can obtain an approximate $1-\alpha$
confidence interval for $g(u_0)$.

\section{Concluding remarks}

 \hskip\parindent
In this paper we have investigated the inference of
single-index mixed models with longitudinal data. We use local
linear regression smoothing to estimate the link function, and
use the generalized estimating equations to estimate the
parametric components. We also construct the estimators of
the variance components. The proposed method avoids the need for
multivariate distribution by only assuming a functional form on
the marginal distribution for each time. The covariance structure
across time is treated as a nuisance. A key feature of our
approach is that we transform a restricted least squares problem
to an unrestricted least squares problem by solving the estimating
equations to estimate the parametric components. The asymptotic
variance of our estimator for parametric components is the same as
that obtained by Wang et al. (2010) in pure single-index models.

In longitudinal studies, sometimes the covariance structure is
very complex; that is, the covariance matrix of outcome variable
may be of a general form, allowing $V$ to have $\frac{1}{2}m(m-1)$
parameters. Our method can be extended to study this type of problem. In
particular, the estimators obtained using our method will be
efficient only if the observations on a subject are independent. The
estimating equations described in this paper can be considered as
an extension of the quasi-likelihood to the case where the second
moment cannot be fully specified in terms of the expectation but
rather additional correlation parameters must be estimated. It is
the independence across subjects that allows us to consistently
estimate these nuisance parameters where this could not be done
otherwise.


\


\renewcommand{\baselinestretch}{1.3}
\begin{center}
{REFERENCES }
\end{center}

\begin{description}
\renewcommand{\baselinestretch}{1.3}

\item Arnold, S. F. (1981). {\it The Theory of Linear Models and Multivariate Analysis}. John Wiley \& Sons, New York.

\item Bai, Y., Fung, W. K. and Zhu, Z. Y. (2009). Penalized
quadratic inference functions for single-index models with
longitudinal data. {\it J. Multivariate Anal.} \textbf{100}
152--161.

\item Cai, T., Cheng, S. C. and Wei, L.J. (2002). Semiparametric
mixed-effects models for clustered failure time data. {\it J.
Amer. Statist. Assoc.} \textbf{97} 514--522.

\item Carro, J. M. (2007). Estimating dynamic panel data discrete
choice models with fixed effects. {\it J. Econometrics}
\textbf{140} 503�-528.

\item Carroll, R. J., Fan, J. Gijbels, I. and Wand, M. P. (1997).
Generalized partially linear single-index models. {\it J. Amer.
Statist. Assoc.} \textbf{92} 477--489.

\item Craven, P. and Wahba, G. (1979), Smoothing and noisy data
with spline functions: estimating the correct degree of smoothing
by the method of generalized cross-validation, {\it Numer. Math.}
\textbf{31} 377-403.

\item Doukhan, P., Massart, P. and Rio, E. (1995). Invariance
principles for absolutely regular empirical processes. {\it Ann.
Inst. H. Poincar\'{e} Prob. Statist.} \textbf{31} 393-427.

\item Fan, J. and Gijbels, I. (1996). {\it Local Polynomial
Modeling and Its Applications}. Chapman and Hall, London.

\item Field, C. A., Pang, Z. and Welsh, A. H. (2008). Bootstrapping Data with Multiple Levels of Variation. {\it Canad. J. Statist.} \textbf{36} 521--539.



\item Gu, C. and Ma, P. (2005). Optimal smoothing in nonparametric
mixed-effect models. {\it Ann. Statist.} \textbf{33} 1357--1379.

\item Hall, P. (1989). On projection pursuit regression. {\it Ann.
Statist.} \textbf{17} 573--588.

\item Hall, P. and Maiti, T. (2006). Nonparametric estimation of
mean-squared prediction error in nested-error regression models.
{\it Ann. Statist.} \textbf{34} 1733--1750.

\item H\"ardle, W., Hall, P. and Ichimura, H. (1993). Optimal
smoothing in single-index models. {\it Ann. Statist.} \textbf{21}
157--178.

\item Honor�, B. and Kyriazidou, E. (2000). Panel data discrete
choice models with lagged dependent variables. {\it Econometrica}
\textbf{68} 839�-874.

\item Hristache, M., Juditsky, A. and Spokoiny, V. (2001). Direct
estimation of the index coefficient in a single-index model. {\it
Ann. Statist.} \textbf{29} 595--623.

\item Hsing, T. and Carroll, R. J. (1992). An asymptotic theory
for sliced inverse regression. {\it Ann. Statist.} \textbf{20}
1040--1061.

\item Ichimura, H. (1993). Semiparametric least squares(SLS) and
weighted SLS estimation of single-index models. {\it J.
Econometrics} \textbf{58} 71--120.

\item Jiang, J. (1998). Consistent estimators in generalized
linear mixed models. {\it J. Amer. Statist. Assoc.} \textbf{93}
720--729.

\item Jiang, J. (1999). Conditional inference about generalized
linear mixed models. {\it Ann. Statist.} \textbf{27} 1974--2007.

\item Jiang, J. (2006). {\it Linear and Generalized Linear Mixed
Models and Their Applications.} Springer.

\item Jiang, J., Jia, H., and Chen, H. (2001). Maximum posterior
estimation of random effects in generalized linear mixed models.
{\it Statistica Sinica} \textbf{11} 97�-120.

\item Jiang, J. and Zhang, W. (2001). Robust estimation in
generalized linear mixed models. {\it Biometrika} \textbf{88}
753�-765.

\item Ke, C. L. and Wang, Y. D. (2001). Semiparametric nonlinear
mixed-effects models and their applications. {\it J. Amer.
Statist. Assoc.} \textbf{96} 1272--1281.



\item Li, K. C. (1991). Sliced inverse regression for dimension
reduction (with discussion). {\it J. Amer. Statist. Assoc.}
\textbf{86} 316--342.


\item Liang, H., Wu, H. L. and Carroll, R. J. (2003). The
relationship between virologic and immunologic responses in AIDS
clinical research using mixed-effects varying-coefficient models
with measurement error. {\it Biostatistics} \textbf{4} 297--312.

\item Liang, K. Y. and Zeger, S. L. (1986). Longitudinal data analysis using generalized linear models. {\it Biometrika} \textbf{73} 13--22.

\item Lin, X. H. \& Carroll, R. J. (2000). Nonparametric function
estimation for clustered data when the predictor is measured
without/with error. {\it J. Amer. Statist. Assoc.} \textbf{95},
520--534.

\item Lo\`{e}ve, M. (2000). {\it Probability Theory I}, 4th
Edition. Springer-Verlag.

\item Mack, Y. P. and Silverman, B. W. (1982). Weak and strong
uniform consistency of kernel regression estimates. {\it Z.
Wahrsch. verw. Gebiete} \textbf{61} 405--415.

\item Masry, E. and Tj\o stheim, D. (1995). Nonparametric
estimation and identification of nonlinear ARCH time series:
Strong convergence and asymptotic normality. {\it Econometric
Theory} \textbf{11} 258--289.

\item Naik, P. and Tsai, C. L. (2000). Partial least squares
estimator for single-index. {\it J. Roy. Statist. Soc. ser. B}
\textbf{62} 763--771.


\item Ruckstuhl, A. F., Welsh, A. H., and Carroll, R. J. (2000).
Nonparametric function estimation of the relationship between two
repeatedly measured variables. {\it Statistica Sinica }
\textbf{10} 51--71.

\item Stute,W. and Zhu, L. X. (2005). Nonparametric checks for
single-index models. {\it Ann. Statist.} \textbf{33} 1048--1083.

\item Thall, P. and Vail, S. C. (1990). Some covariance models for
longitudinal count data with over dispersion, {\it Biometrics}
\textbf{46} 657�-671.

\item Wang, J. L., Xue, L. G., Zhu, L. X. and Chong, Y. S. (2010).
Partial-linear single-index model with noised variable. {\it Ann.
Statist.} \textbf{38} 246--272.

\item Wang, Y. G., Lin, X. and Zhu, M. (2005). Robust estimating
functions and bias correction for longitudinal data analysis,
{Biometrics} \textbf{61} 684�-691.

\item Wu, H. L. and Zhang, J. T. (2002). Local polynomial
mixed-effects models for longitudinal data. {\it J. Amer. Statist.
Assoc.} \textbf{97} 883--897.

\item Xia, Y. (2006). Asymptotic distributions for two estimators
of the single-index model. {\it Econometric Theory} \textbf{22}
1112--1137.

\item Xia, Y. and H\"ardle, W. (2006). Semi-parametric estimation
of partially linear single-index models. {\it J. Multi. Anal.}
\textbf{97} 1162 - 1184.

\item Xia, Y. and Li, W. K. (1999). On single-index coefficient
regression models. {\it J. Amer. Statist. Assoc.} \textbf{94}
1275--1285.

\item Xia, Y., Tong, H. Li, W. K. and Zhu L. X. (2002). An
adaptive estimation of dimension reduction space. {\it J. R.
Statist. Soc.} B \textbf{64} 363--410.

\item Xue, L. G. (2010). Empirical likelihood local polynomial
regression analysis of clustered data. {\it J. Scandinavain
Statistics}, to be appeared.

\item Xue, L. G. and Zhu L. X. (2006). Empirical likelihood for
single-index models. {\it J. Multivariate Anal.} \textbf{97}
1295--1312.

\item Yu, Y. and Ruppert, D. (2002). Penalized spline estimation
for partially linear single-index models. {\it J. Amer. Statist.
Assoc.} \textbf{97} 1042--1054.

\item Zeger, S. L. and Diggle, P. J. (1994). Semiparametric Models
for Longitudinal Data with Application to CD4 Cell Numbers in HIV
Seroconverters. {\it Biometrics} \textbf{50} 689--699.

\item Zhang, D. W. and Lin, X. H. (2003). Hypothesis testing in
semiparametric additive mixed models. {\it Biostatistics}
\textbf{4} 57--74.

\item Zhang, D., Lin, X. H., Raz, J., et al. (1998).
Semiparametric stochastic mixed models for longitudinal data.
{\it J. Amer. Statist. Assoc.} \textbf{93} 710--719.


\item Zhu, L. X. and Ng, K. W. (1995). Asymptotics for Sliced
Inverse Regression. {\it Statistica Sinica} \textbf{5} 727-736.

\item Zhu L. X. and Xue L. G. (2006). Empirical likelihood
confidence regions in a partially linear single-index model. {\it
J. Roy. Statist. Soc. ser. B} \textbf{68} 549--570.

\end{description}

\newpage

\end{document}